\documentclass[twocolumn,showpacs,preprintnumbers,aps,amsmath,amssymb]{revtex4}

\pdfoutput=1

\usepackage{amsmath}
\usepackage{stmaryrd}
\usepackage{txfonts}
\usepackage{amssymb}
\usepackage{mathrsfs}
\usepackage{graphicx}
\usepackage{dcolumn}
\usepackage{bm}
\usepackage{epsfig}
\usepackage{color}
\usepackage[colorlinks=true, linkcolor=blue, citecolor=blue, urlcolor=blue]{hyperref} 
\usepackage{epstopdf}

\begin{document}      

\title{Revealing the Hidden Structural Phases of FeRh} 

\author{Jinwoong Kim}
\email{jinwoong.kim@csun.edu}
\affiliation{Department of Physics, California State University, Northridge, CA 91330, USA}
\author{R. Ramesh}
\affiliation{Department of Materials Science, University of California, Berkeley, California 94720, USA}
\author{Nicholas Kioussis}
\email{nick.kioussis@csun.edu}
\affiliation{Department of Physics, California State University, Northridge, CA 91330, USA}

\date{\today}      

\begin{abstract}
{\it Ab initio} electronic structure calculations reveal that tetragonal distortion has a dramatic effect on the relative stability of the various magnetic structures (C-, A-, G-, A$'$-AFM, and FM) of FeRh giving rise to a wide range of novel stable/metastable structures and magnetic phase transitions between these states. We predict that the {\it cubic} G-AFM structure, which was believed thus far to be the ground state, is metastable and that the {\it tetragonally} expanded G-AFM is the stable structure. The low energy barrier separating these states suggests phase coexistence at room temperature. We propose a novel A$'$-AFM phase to be the {\it global} ground state among all magnetic phases which arises from the strain-induced tuning of the exchange interactions.
The results elucidate the underlying mechanism for the recent experimental findings of electric-field control of magnetic phase transition driven via tetragonal strain.
The novel magnetic phase transitions open interesting prospects for exploiting strain engineering for the next-generation memory devices.
\end{abstract}

 
\maketitle      
The binary FeRh metallic alloys continue to be the subject of intense experimental and theoretical research due to the wide range
of fascinating magnetic and transport properties and their potential applications in thermally assisted magnetic recording media ~\cite{Fullerton03},
magnetic cooling~\cite{Thiele2011}, ultrafast (ps) switching~\cite{Ju2004}, and
room-temperature antiferromagnetic memory resistor~\cite{Marti14}.

\begin{figure*}[ht]
	\centering
	\includegraphics[width=18cm,natwidth=625,natheight=270]{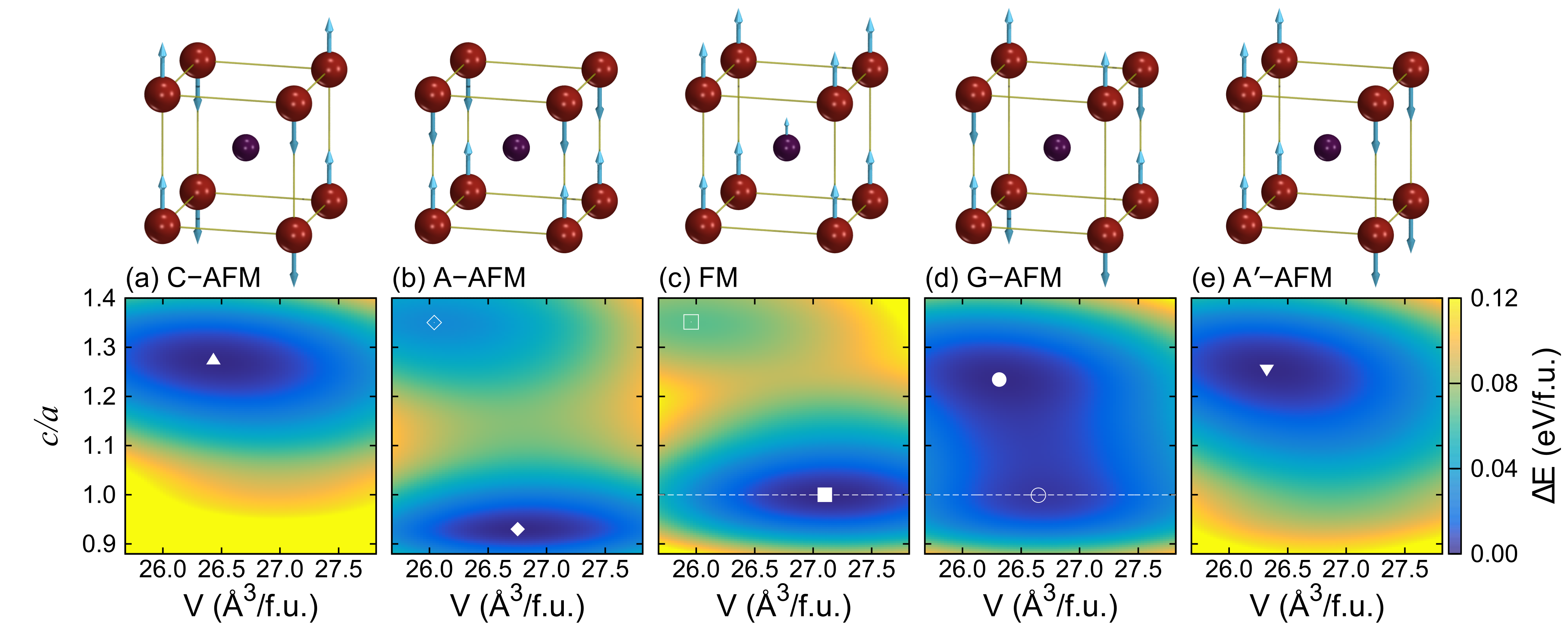}
	\caption{(Color online) Contour plots of relative energy, $\Delta E_i=E_i -E_{i}^{\text{min}}$, in the $c/a$ versus volume parameter space for the $i^{th}$ = C-AFM, A-AFM, FM, G-AFM, A$'$-AFM phase of FeRh
		shown in the upper horizontal panels, respectively. Here,
		the large/bright (small/dark) spheres denote Fe (Rh) atoms, the arrows the local magnetic moments, and $E_{i}^{\text{min}}$ is the global energy minimum for the $i^{th}$ phase denoted by closed symbols, while open symbols denote local energy minima.}
	\label{Fig1}
\end{figure*}

The near equiatomic bulk FeRh can exhibit a chemically
ordered bcc-B2 (CsCl-type) structure or a chemically disordered
fcc $\gamma$ structure at room temperature. The fcc phase is nonmagnetic~\cite{Lommel,Miuajima} while the bcc phase undergoes an unusual first-order isostructural magnetic phase transition
from a G-type antiferromagnetic (G-AFM) phase to a ferromagnetic (FM) phase when heated above $\sim$ 370 K~\cite{Fallot1938,Fallot1939,Kouvel1962}.
Neutron scattering~\cite{Shirane1964} and x-ray magnetic circular dichroism~\cite{Stamm2008} experiments
report that the low-temperature G-AFM phase  is characterized by a magnetic structure [Fig. \ref{Fig1}(d)] in which the Fe local moments are $\sim \pm$ 3 $\mu_B$
and negligible moments on Rh sites,  while in the high-temperature FM phase [Fig. \ref{Fig1}(c)] the iron and rhodium local moments  are $\sim$ 3.2 $\mu_B$ and $\sim$ 1 $\mu_B$, respectively.
The metamagnetic transition is accompanied by volume expansion of $\sim$ 1\% and a large drop in resistivity indicating coupling between the electronic, magnetic and structural degrees of freedom~\cite{Kouvel1962,Bergevin1961}.
Application of hydrostatic pressure suppresses the FM phase, i.e., under a critical pressure of 60 kbar the system transforms directly
from the G-AFM to the paramagnetic phase~\cite{Zakharov}.
The underlying origin of the G-AFM to FM transformation is controversial and remains unresolved.
Proposed mechanisms include changes in the electronic entropy~\cite{Tu1969},
instability of the Rh magnetic moment~\cite{Gruner2003}, and magnetic excitations~\cite{Mavropoulos2011}.

Most of density functional theory (DFT) calculations ~\cite{Gruner2003,Mavropoulos2011,Moruzzi92} to date have focused on the
electronic structure properties solely of the {\it cubic} (bcc) structure under hydrostatic pressure.
On the other hand, recent experiments on FeRh thin films epitaxially grown on ferroelectric BaTiO$_3$~\cite{Cherifi2014,ZQLiu2016,Bibes2015} or piezoelectric (PMN-PT)~\cite{Ramesh2015} substrates have provided evidence
of an isothermal electric-field control of the magnetic phase transition driven via piezoelectirc biaxial strain. These results raise the intriguing question of the effect
of strain on tuning the interplay between FM and AFM spin correlations and hence the stability of the FeRh phases.
Interestingly, early experiments find that FeRh undergoes a transition from a bcc to a fcc structure under plastic deformation~\cite{Lommel} or uniaxial strain~\cite{Miuajima}.
Subsequent electronic structure calculations~\cite{Pugacheva} confirmed the Bain tetragonal deformation path between the lower-energy AFM {\it fcc} and the FM {\it bcc} structures, but did not consider the G-AFM phase.

In this report first principles electronic structure calculations reveal
that tetragonal distortion gives rise to surprising new effects on the relative stability of the various magnetic structures
(C-, A-, G-, A$'$-AFM, and FM) of FeRh. We predict
that the {\it cubic} G-AFM structure, which was believed thus far to be the ground state, is metastable and that the {\it tetragonally} expanded G-AFM is the ground state. We find a low energy barrier separating these two structures suggesting transition between these
structures at room temperature. More importantly, we demonstrate that a novel A$'$-AFM structure  is the most stable magnetic phase
among all considered phases.

We have employed DFT calculations within the projector-augmented wave method~\cite{Blochl94}, as implemented in the Vienna \textit{ab initio} simulation package~\cite{Kresse96a,Kresse96b,KressePAW}. The generalized gradient approximation is used to describe the exchange-correlation functional as parameterized by Perdew-Burke-Ernzerhof~\cite{PBE}.
The plane-wave cutoff energy is 400 eV and the Brillouin zone is sampled using $22\times11\times11$ \textit{k}-mesh for the A$'$-AFM phase and $15\times15\times11$ \textit{k}-grid for the other phases.
Convergence tests employing a 550 eV cutoff energy and $19\times19\times14$ \textit{k}-grid show
that the total energy is converged to less than 0.3 meV/(f.u.).
Spin-orbit coupling is not included and all results are for collinear systems ~\cite{SOC}.
The two dimensional $c/a$ versus volume energy map is calculated on $9 (\text{volume})\times14 (c/a)$ grid mesh and interpolated using a cubic spline interpolation. We have double checked the existence of metamagnetic structures via individual structural relaxation with convergence criteria of $\Delta E < 10^{-8} $ eV.
Due to the broken $C_4$ rotational symmetry along [001] in the A$'$-AFM phase (associated with its magnetic configuration) one would expect that the orthorhombic structure may be more stable than the tetragonal one. Nevertheless, we find that even after structural relaxation the stable A$'$-AFM phase preserves its tetragonal symmetry with 0.5 \% in-plane lattice constant difference.
For the phonon calculations we have employed the VASP and PHONOPY ~\cite{phonopy} codes and a 16 atom-supercell to determine the dynamical matrix with a cutoff energy of 500 eV and $12\times12\times12$ \textit{k}-mesh.

\textit{Volume versus $c/a$ Phase Diagram---}
The upper panels in Fig. \ref{Fig1} show the C-AFM, A-AFM, FM, G-AFM, and A$'$-AFM magnetic structures considered in this work.
The AFM ordering can generally be described as alternating FM planes which are antiferromagnetically coupled along the sheet normal direction, where the FM planes for the cubic C-, A-, G-, and A$'$-AFM, structures are the (110), (001), (111), and (100) planes, respectively. In contrast to all
the AFM phases chracterized by a  negligible Rh local moment, in the FM phase the non-zero Rh local moment is induced by the
non-vanishing net exchange field from the nearest-neighbor Fe atoms~\cite{Mavropoulos2011}.

The lower panels in Fig. \ref{Fig1} show contour plots of the relative energy landscape, $\Delta E_i=E_i -E_{i}^{\text{min}}$, of the five magnetic structures ($i$=C-, A-, G-, A$'$-AFM, FM) as a function of volume and tetragonal distortion, $c/a$ ratio. Here, $E_{i}^{\text{min}}$ is the global energy minimum (denoted by closed symbol) of the $i^{th}$ structure, while the open symbols denote local energy minima.
The calculations reveal that tetragonal distortion gives rise to a wide range of novel stable magnetic structures.
As expected, along the $c/a$=1 line (i.e. cubic phase) both G-AFM and FM
phases with equilibrium lattice constants (see Table I) of 2.987 $\text{\AA}$ and 3.004 $\text{\AA}$, respectively,
exhibit substantial energy minima.
Interestingly, we find additional stable structures away from the $c/a$=1 line in {\it all} considered magnetic phases. First, we demonstrate that in contrast to previous experimental and theoretical consensus, the {\it cubic} ($c/a$=1) G-AFM is not the ground state. Rather the tetragonally distorted phase with $c/a$=1.235 is the {\it global} energy minimum. Second and more importantly, we predict that tetragonal distortion renders the A$'$-AFM phase [Fig. \ref{Fig1}(e)] with $c/a$=1.257 to be the {\it global} ground state among all magnetic phases.
Third, the global energy minimum of the C-AFM (A-AFM) phase occurs at $c/a$ = 1.273 (0.930).
The A-AFM and FM phases also exhibit additional metastable tetragonal structures with $c/a = 1.350$ and $1.352$, respectively, which however,
are not the global energy minimum in contrast to the G-AFM phase. Thus, only the G-AFM and FM phases are metastable/stable under cubic symmetry while the tetragonal structures are commonly found in diverse magnetic structures.

\begin{figure}
	\includegraphics[width=8.6cm,natwidth=2025,natheight=1508]{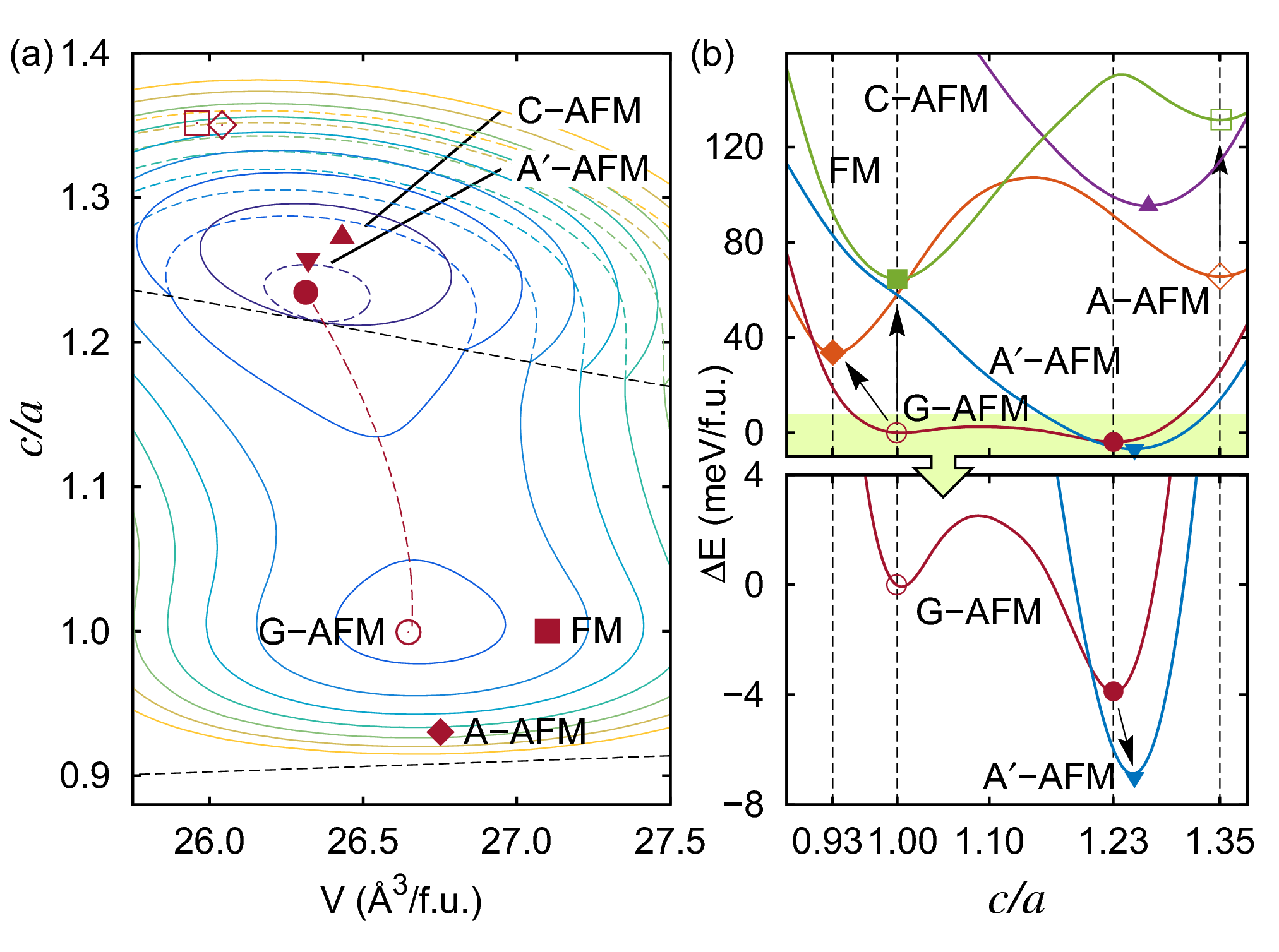}
	\caption{(Color online) (a) Energy contours of most stable magnetic phases in the volume versus $c/a$ parameter space, which is divided in three regions
		by the dashed lines around $c/a$ = 1.2 and 0.9, respectively. Solid contours in the upper (middle) regions correspond to the stable A$'$-AFM (G-AFM) phases while
		dashed contours in the upper region correspond to the metastable G-AFM phase.
		Closed (open) symbols denote the global (local) energy minima for each phase and the
		red dashed curve denotes the Bain path for the G-AFM phase.
		(b) Energies relative to that of bcc G-AFM versus $c/a$, where the curves are the minimum energy paths over volume at a given $c/a$ ratio. Arrows denote new magnetic phase transitions.}
	\label{Fig2}
\end{figure}

Fig. \ref{Fig2}(a) displays equal energy contours of the most stable magnetic structures in the volume versus $c/a$ parameter space,
which is divided in three regions by the dashed lines around $c/a$ = 1.2 and 0.9, respectively. In the upper (lower) regions
the A$'$-AFM (A-AFM) phase is more stable, while in the middle region bounded by the two lines the G-AFM has the lowest energy.
The solid curves in the upper and middle regions denote the equal energy contours of the stable A$'$-AFM and G-AFM phases. We also
show for comparison with dashed curves in the upper region the equal energy contours of the metastable G-AFM phase. The closed (open) symbols denote the global (local) energy minima of the stable (metastable) structures for each magnetic phase.
One can see that the global energy minima of the G-, C- and A$'$-AFM phases occur close to $c/a\thickapprox$1.26 and $V\thickapprox26.3$ \AA$^3$.

\begin{table}[b]
	\centering
	\caption{Calculated equilibrium lattice constants in \AA, $c/a$ ratio, total energy per f.u. in meV relative to that of the bcc G-AFM phase, and magnetic moments of Fe and Rh in $\mu_{B}$ for the
		stable magnetic structures of FeRh. Experimental values are listed in parentheses for comparison.
		$^{*}$ The low Rh moment in the C-AFM phase disappears after orthorhombic relaxation in which the in-plane lattice difference is smaller than 0.2 $\%$.}
	\begin{ruledtabular}
		\begin{tabular}{ c | c c c r c c } 
			Phase &  $a$      &   $c$   &  $c/a$  & $\Delta E$ & $|\text{M}_{\text{Fe}}|$ & $|\text{M}_{\text{Rh}}|$ \\ 
			\hline
			C-AFM & 2.749     & 3.498   & 1.27    &  95.4      & 2.95      & 0.12$^{*}$ \\ [0.6ex] 
			A-AFM & 3.064     & 2.850   & 0.93    &  33.7      & 3.12      & -          \\ 
			& 2.682     & 3.621   & 1.35    &  65.8      & 2.66      & -          \\ [0.6ex] 
			FM    & 3.004     & 3.004   & 1.00    & 64.5       & 3.17      & 1.02       \\ 
			& 2.678     & 3.620   & 1.35    & 131.6      & 2.54      & 0.15       \\ [0.6ex] 
			G-AFM & 2.987     & 2.986   & 1.00    &  0.0       & 3.12      & -          \\ 
			& (2.986)\footnotemark[1]   & (2.986)\footnotemark[1] & (1.00)\footnotemark[1]  & - & (3.3)\footnotemark[1] \\ 
			& 2.773     & 3.423   & 1.23    & -3.9       & 2.90      & -          \\ 
			& (2.81)\footnotemark[2]    & (3.35)\footnotemark[2]  & (1.19)\footnotemark[2]  \\ 
			& (2.83)\footnotemark[3]    & (3.33)\footnotemark[3]  & (1.18)\footnotemark[3]  \\ [0.6ex] 
			A$'$-AFM & 2.761  & 3.472   & 1.26    & -7.0       & 3.15      & -          \\ 
		\end{tabular}
	\end{ruledtabular}
	\label{tab:my_label}
	\footnotetext[1]{Reference~\cite{Shirane1964}}
	\footnotetext[2]{Reference~\cite{Miuajima}}
	\footnotetext[3]{Reference~\cite{Takahashi}}
\end{table}

Fig. \ref{Fig2}(b) shows the variation of the total energy per formula unit relative to that of the {\it cubic} G-AFM phase versus $c/a$ along the tetragonal Bain path for the A-, C-, A$'$-, G-AFM and FM phases. For example the Bain path (where $\Delta c/\Delta a$ is constant) for the G-AFM phase
connecting the stable (closed circle) and metastable (open circle) states in Fig. \ref{Fig2}(a) is displayed with the red dashed curve.
There are similar minimum energy Bain trajectories [not shown in Fig. \ref{Fig2}(a)] connecting the closed and open symbols for the other phases.
The relative energy of the bcc FM phase with respect the bcc G-AFM is 64.5 meV/f.u. in a good agreement with previous GGA results ~\cite{Gruner2003}, which is twice the first order transition temperature  of $\sim$ 30 meV. Under tetragonal tensile strain the G-AFM exhibits a stable phase at $c/a$ = 1.23 which is 3.9 meV/f.u. lower in energy than that of the bcc G-AFM structure.
Our result disagrees with that of recent {\it ab initio} electronic structure calculations reporting that the cubic G-AFM phase is the ground state ~\cite{Aschauer}.
The energy barrier ongoing from the tetragonal to the bcc phase is 6.4 meV/f.u. indicating co-existence of the two phases above
80 K. The fact that the FeRh is usually synthesized at temperatures higher than 80 K might explain why the tetragonal G-AFM structure has not been observed experimentally.
It is important to note that  the {\it cubic} G-AFM phase is softer under tetragonal deformation compared to the FM phase leading to an increase of
the relative energy, $\Delta E=E_{\text{FM}}-E_{\text{G-AFM}}$ under tensile strain. This is exactly the underlying mechanism of recent experimental findings~\cite{Cherifi2014,ZQLiu2016,Bibes2015} that the metamagnetic transition temperature
of epitaxial FeRh films on ferroelectric substrate can be controlled by applied voltage through the tetragonal strain that introduces huge magnetoelectric coupling.
Furthermore, the predicted stable tetragonal G-AFM phase with $c/a$ = 1.235 is consistent with earlier experimental findings that plastic deformation~\cite{Lommel} or uniaxial strain~\cite{Miuajima}
induces a transition from the bcc G-AFM to the fcc L1$_0$ structure ($c/a = \sqrt{2}$), where the body-centered tetragonal (bct) with $c/a$=$1.19$ and the fcc phases coexist~\cite{Miuajima}.
Subsequent experiments~\cite{Takahashi} have shown that  the bct structure is independent from the fcc, rather than an intermediate phase between the bcc and fcc structures.

The calculated equilibrium structural parameters, relative energy per f.u., and magnetic moments of the various stable and metastable magnetic phases are summarized in Table I and compared with experiment. The bcc FM phase exhibits the largest Fe magnetic moment which however decreases in the
metastable bct FM phase. Similarly, the tetragonally expanded A- and G-AFM phases show lower Fe local magnetic moment. It is interesting to note
that the larger Fe magnetic moment (3.15 $\mu_{B}$) in the bct A$'$-AFM phase compared to that (2.90 $\mu_{B}$) in the bct G-AFM allows
distinguishing the two phases experimentally by measuring the hyperfine field.

\textit{Strain-induced Novel Phase Transitions---}
The results in Fig.\ref{Fig2}(a) and (b) suggest that tetragonal distortion can induce several potential novel magnetic phase transitions in addition to the well-known temperature-driven bcc G-AFM to bcc FM
transition where entropy renders the FM phase stable at higher temperatures.
First, tetragonal deformation can induce an irreversible magnetic phase transition from the bct G-AFM to bct A$'$-AFM phase both of which have $c/a\thicksim$ 1.23. The bct A$'$-AFM is a
magnetically protected crystal symmetry since is stable only in the tetragonal structure and cannot undergo a transition to cubic phase even at elevated temperatures. A second plausible
magnetic phase transition is from the bcc G-AFM phase to the compressed ($c/a$=0.93) bct A-AFM phase whose energy relative to the bcc G-AFM phase
is 33.7 meV/f.u. (Table I). This is lower than the corresponding energy of the bcc FM structure indicating that the A-AFM phase can be stabilized by enthalpy (i.e., external uniaxial stress)  rather
than entropy. In fact for thin FeRh films the ferromagnetic ordering of the (001) surfaces of the A-AFM state [Fig.\ref{Fig2}(b)] will facilitate this transition.
A third strain-induced phase transition is from the bct A-AFM to the bct FM phase (both with $c/a\thicksim$1.35) with an energy difference of 65.79 meV/f.u. which is comparable to
that between the bcc FM and bcc G-AFM phases and can be driven via the entropy mechanism around 370 K.

\begin{figure} [t]
	\includegraphics[width=8.6cm,natwidth=2031,natheight=1039]{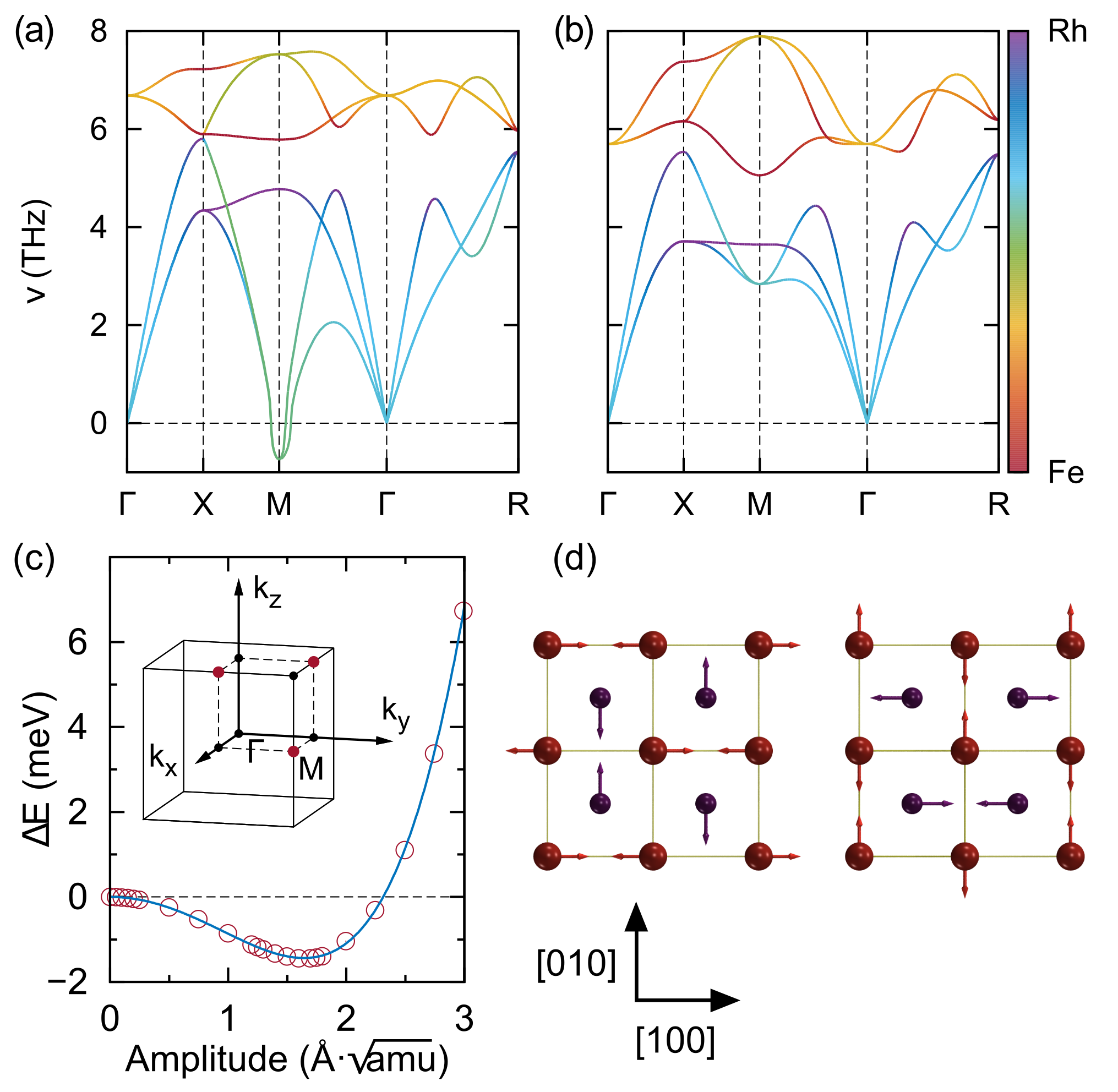}
	\caption{(Color online) Calculated phonon dispersions of (a) bcc G-AFM and (b) bcc FM phases. The colors denote the projected phonon modes on the atomic basis. (c) Energy  of bcc G-AFM FeRh versus phonon amplitude, $\left(\sum_{i}{m_i x_{i}^{2}}\right)^{1/2}$, corresponding to the phonon instability at the \emph{M} point, where $m_i$ and $x_i$ is the mass and displacement of the $i$-th atom, respectively. Circles (solid curves) are the calculated (fitted) energies. (d) Atomic displacement vectors of two degenerate modes at \emph{M} projected on the (001) plane.}
	\label{Fig3}
\end{figure}

\textit{Phonon Instability---}
In order to corroborate the dynamic stability of the bcc G-AFM, bcc FM, bct G-AFM, and bct A$'$-AFM phases we have carried out phonon calculations (phonon dispersions of the two bct structures are not shown here).
The phonon dispersions of the bcc G-AFM and bcc FM structures in Figs. 3 (a) and (b), respectively, show that the low-frequency acoustical branches are associated with the displacements of both Fe and Rh atoms while the optical  (high frequency acoustical) branches are primarily associated with Fe (Rh) displacements.
Surprisingly we find an imaginary frequency at the \emph{M} point for the bcc G-AFM phase implying that it is not a stable structure, in agreement with recent studies~\cite{Aschauer,Wolloch}. Fig. \ref{Fig3}(c) shows the energy landscape versus the phonon amplitude corresponding to the phonon instability at \emph{M} point, where the low energy barrier of 1.5 meV per mode (16 atoms) indicates that the instability will appear at very low temperatures ($< 20$ K).
The structural distortion corresponding to the superposition of the two instability modes shown in Fig. \ref{Fig3}(d) was recently reported in FeRh/W/MgO thin film structures where the tungsten substrate exerts large epitaxial tensile strain (6 \%) facilitating thus the instability at ambient conditions~\cite{Witte}. The instabilities at the other two \emph{M} points presumably disappear due to the large out-of-plane structural suppression.

In summary, we predict that tetragonal distortion, which is ubiquitous in thin FeRh films epitaxially grown on ferroelectric or piezoelectric substrates, can give rise to a wide range of novel stable/metastable magnetic phases and magnetic phase transitions between these states. In contrast to previous experimental and theoretical findings that the cubic G-AFM is the ground state we demonstrate that biaxial strain renders the bct G-AFM to be the stable phase. More importantly the calculations reveal that the novel bct A$'$-AFM phase is the global ground state and that FeRh can undergo an irreversible bct G-AFM to bct A$'$-AFM phase transition.
These results suggest that the experimentally observed cubic G-AFM phase in thin FeRh films may not be uniform, rather it may consist of different AFM domains depending on the local deformation from the substrate lattice mismatch and the external or chemical strain.
We hope that these predictions will rekindle interest in search for new strain-induced metastable phases in other magnetic materials.

The work at CSUN is supported by NSF (USA) Partnership in Research and Education in Materials Research Grant No. 1205734, NSF Nanosystems Engineering Research Center (ERC) for Translational Applications of Nanoscale Multiferroic Systems (TANMS)
Grant No. 1160504 and NSF Grant No. DMR-1532249.


\begin{thebibliography}{34}
\bibitem{SOC} We have checked the effect of spin-orbit coupling (SOC) on the total energies of all the stable/metastable structures. We find that the trend of relative stability remains robust in the presence of SOC. Hence, SOC is not included and all results reported here are for collinear systems.
\end{thebibliography}
\end{document}